\documentclass[pre,preprint,showpacs,eqsecnum]{revtex4}
\begin{document}
\title{Domain statistics in a finite Ising chain}
\author{S.~I.~Denisov}
\email{denisov@sumdu.edu.ua} \affiliation{Department of Mechanics
and Mathematics, Sumy State University, 2, Rimskiy-Korsakov
Street, 40007 Sumy, Ukraine}
\author{Peter~H\"{a}nggi}
\email{hanggi@physik.uni-augsburg.de} \affiliation{Institut
f\"{u}r Physik, Universit\"{a}t Augsburg,
Universit\"{a}tsstra{\ss}e 1, D-86135 Augsburg, Germany}

%\date{Received 8 February 2005}

\begin{abstract}
We present a comprehensive study for the statistical properties of
random variables that describe the domain structure of a finite
Ising chain with nearest-neighbor exchange interactions and free
boundary conditions. By use of extensive combinatorics we succeed in
obtaining the one-variable probability functions for (i) the number
of domain walls, (ii) the number of up domains, and (iii) the number
of spins in an up domain. The corresponding averages and variances
of these probability distributions are calculated and the limiting
case of an infinite chain is considered. Analyzing the averages and
the transition time between differing chain states at low
temperatures, we also introduce a criterion of the
ferromagnetic-like behavior of a finite Ising chain. The results can
be used to characterize magnetism in monatomic metal wires and
atomic-scale memory devices.
\end{abstract}
\maketitle

\section{INTRODUCTION}

The Ising model, pioneered just 80 years ago \cite{Ising}, has
become one of the most popular and useful models of statistical
physics. This model system itself and its numerous generalizations
found wide application for the investigation of not only physical
but also for biological, economical, and social systems, to name
only a few. The model has also been widely used to characterize the
cooperative behaviors in these and other systems. The salient
advantages of the Ising model are that it is generic for systems
with phase transitions, it is very convenient to use, and, moreover,
for  particular cases it can be solved exactly, i.e., its partition
function can be calculated, at least in the thermodynamic limit,
without approximations. Because exact solutions were found only for
a certain one- and two-dimensional versions of the Ising model
\cite{LM,B}, their role for statistical physics is most important.

The ordinary one-dimensional Ising model, which is represented by an
infinite chain of Ising spins, i.e., spins that can either be up or
down, and which do interact with each other via the nearest-neighbor
exchange interaction, does not exhibit the ferromagnetic phase
transition at nonzero temperatures \cite{Ising}. This well-known
result corroborates the known argument of Landau and Lifshitz
\cite{LL}, according to which a long-range order in
\textit{infinite} one-dimensional systems with \textit{short-range}
interactions is absent. The problem of long-range ordering, which
can emerge in such systems when these conditions are violated, is of
prominent theoretical importance. Its solution for infinite Ising
chains with \textit{long-range} interactions between spins has been
the subject of a number of remarkable studies (see, e.g.,
Refs.~\cite{R,D,T,SS,FS,ACCN,WG}).

A priori, the statistical mechanics of a \textit{finite} Ising chain
with only exchange (i.e., short-range) interaction seems not to
present an interesting topic. This is so, because it does not
exhibit macroscopic ferromagnetic order. A detailed investigation of
this model is important, however, by the following motivating
reasons. First, the domain statistics in such  finite chains, i.e.,
a probability description of forming domains, domain lengths, and
domain walls contains most valuable information on the thermal
equilibrium state. To the best of our knowledge, these statistics
have not been studied before. The main problem is that the domain
characteristics are not ordinary thermodynamic quantities, i.e.,
they are not readily expressed through the partition function. In
short, there are no conventional methods to extract them. Second, a
finite Ising chain represents an appropriate phenomenological model
for describing magnetism in \textit{monatomic} metal wires deposited
on substrates. Indeed, as it has been discovered experimentally
\cite{G}, a Co chain on Pt substrate is characterized by the
exchange coupling (this justifies the nearest-neighbor
approximation), very large magnetic anisotropy (this justifies the
approximation of atomic magnetic moments by Ising spins), and
long-range ferromagnetic order (this justifies the use of finite
Ising chains). We do emphasize that in contrast to the case with
infinite Ising chains for which ferromagnetic order is forbidden at
all nonzero temperatures \cite{NM},  finite chains can exhibit the
ferromagnetic-like behavior (see also Sec.~IV). Finally, it is very
likely that the one-dimensional magnets, which are modelled by
finite Ising chains, will have important implication for magnetic
data-storage technology \cite{K}.

In this paper, using a variety of combinatorial approaches, we
investigate thoroughly the domain statistics in a finite Ising
chain. In Sec.~II, we describe the model and introduce the main
definitions. The joint probability functions that describe the
domain structure of a chain are calculated in Sec.~III by the
combinatorial method. In Sec.~IV, we demonstrate that the number of
domain walls in a finite Ising chain is binomially distributed. We
then introduce a criterion of its ferromagnetic-like behavior and
consider the  limiting case of an infinite chain. In Sec.~V, we
derive the probability function for the number of up domains and
calculate its average and its variance. The probability function for
the number of spins in an up domain (i.e., domain length) and its
numerical characteristics are determined in Sec.~VI. We summarize
our novel findings  in Sec.~VII. Some of our technical details and
manipulations are deferred to the Appendices.

\section{MODEL AND NOTATIONS}

We consider a finite Ising chain with free boundary conditions that
contains an even number $N$ of Ising spins. We assume that the spins
interact through the nearest-neighbor ferromagnetic coupling $J(>0)$
and the spin variables $\sigma_{i}$ ($i=1,\dots,N$) assume only two
values $+1$ and $-1$, respectively,  corresponding to the up and
down spin orientations. For a given spin configuration $\{
\sigma_{i} \}$, the chain energy is written in the form
\begin{equation}
    E_{N}(\{\sigma_{i}\}) = -J\sum_{i=1}^{N-1}\sigma_{i}
    \sigma_{i+1}.
    \label{eq:energy1}
\end{equation}
According to the Gibbs distribution, the probability of this
configuration is given by
\begin{equation}
    W_{N}(\{\sigma_{i}\}) = \frac{1}{Z_{N}}e^{-\beta
    E_{N}(\{\sigma_{i}\})},
    \label{eq:prob1}
\end{equation}
where $\beta$ denotes the inverse temperature measured in energy
units, and $Z_{N} = \sum_{\{\sigma_{i}\}} \exp[-\beta E_{N}(\{
\sigma_{i} \} )]$ is the partition function of a chain. Using,
e.g., the transfer matrix method \cite{KW}, $Z_{N}$  can be
evaluated exactly, yielding the well-known result
\begin{equation}
    Z_{N} = 2^{N} \cosh^{N-1} (\beta J).
    \label{eq:partition1}
\end{equation}

In order to characterize the domain distribution in a chain, we
introduce the number of up spins, $s$, the number of up domains,
$p$, the number of domain walls, i.e., the number of up-down and
down-up spin pairs, $k$, and the number of spins in the first
up-domain, $l$. These numbers satisfy the conditions: $0 \leq s \leq
N$, $0 \leq p \leq N/2$, $0 \leq k \leq N-1$, $0 \leq l \leq N$.
These numbers are not independent because, for example, if $p=0$
then $s=k=l=0$, and if $p=N/2$ then $s=N/2$, $k=N-1$, and $l=1$. The
introduced quantities are random due to thermal agitation, and our
main objective is to calculate the one-variable probability
functions $P_{N}(k)$, $P_{N}(p)$, and $P_{N}(l)$ that describe in
detail domain statistics in a chain of $N$ Ising spins. (Notice that
some features of the probability function of magnetization have been
studied in \cite{ADR}.) To this end, we also introduce the
four-variable probability function $P_{N}(s,p,k,l)$ representing the
joint probability that a chain is characterized by the parameters
$s$, $p$, $k$, and $l$. Taking into account that
\begin{equation}
    E_{N}(\{\sigma_{i}\}) = E_{N}(k) = -J(N-1)+2Jk
    \label{eq:energy2}
\end{equation}
and
\begin{equation}
    W_{N}(\{\sigma_{i}\}) = W_{N}(k) = \frac{1}{Z_{N}}e^{-\beta
    E_{N}(k)},
    \label{eq:prob2}
\end{equation}
this probability function can be written as
\begin{equation}
    P_{N}(s,p,k,l) = W_{N}(k)K_{N}(s,p,k,l),
    \label{eq:joint4}
\end{equation}
where $K_{N}(s,p,k,l)$ is the number of spin configurations
possessing the same set of the non-negative integer variables $s$,
$p$, $k$, and $l$. In accordance with the basic laws of probability
theory \cite{F}, all the one-variable probability functions can be
determined by fixing one variable and summing $P_{N}(s,p,k,l)$ over
the admissible values of all remaining variables.

\section{JOINT PROBABILITY FUNCTIONS}

\subsection{Number of spin configurations}

The chain states that we describe in terms of the four variables
mentioned above are, in general, degenerate and $K_{N}(s,p,k,l)$
spin configurations correspond to each of those states. The states
with $s=p=k=l=0$ and $s=l=N$, $p=1$, $k=0$ are characterized by only
one spin configuration $\{\sigma_{i}=-1\}$ and $\{\sigma_{i}=+1\}$
for all $i$, respectively. Therefore $K_{N}(0,0,0,0) =
K_{N}(N,1,0,N) = 1$. For counting $K_{N}(s,p,k,l)$ in other cases,
when $1 \leq s \leq N-1$, we use combinatorial methods. Within their
framework, we consider an Ising chain with fixed $s$, $p$, $k$, and
$l$ as an alternate sequence of $p$ up boxes and $k-p+1$ down boxes
in which $s$ up spins and $N-s$ down spins are distributed.

Because the first up box must contain $l$ up spins and in each other
up box must be at least one up spin (hence the condition $s-l \geq
p-1$ must hold), the number $M_{\uparrow}$ of different
distributions of $s$ up spins over $p$ up boxes equals $C_{s-l-1}^
{p-2}$. Here, the binomial coefficients $C_{n}^{m}$ with integers
$n$ and $m$ are defined as follows: $C_{n}^{m} = n!/(n-m)!m!$ if $n
\geq m \geq 0$, $C_{n}^{m} = 0$ if $m > n \geq 0$ or if $n \geq 0$
and $m<0$, and $C_{n}^{0} = C_{n}^{n} = 1$ for all integer $n$.
Using these properties, we can represent $M_{\uparrow}$ in the form
$M_{\uparrow} = C_{s-l-1}^{p-2 + \Delta_{spkl}}$ ($\Delta_{spkl} =
\delta_{s,0} \delta_{p,0} \delta_{k,0} \delta_{l,0}$, $\delta_{n,m}$
is the Kronecker symbol), which is valid for $0 \leq s \leq N$.

Similarly, the number $M_{\downarrow}$ of different distributions of
$N-s$ down spins over $k-p+1$ down boxes, each of which contains at
least one down spin, is given by $M_{\downarrow} = C_{N-s-1}^{k-p}$.
By the same reason as in the previous case, this formula is also
valid for all values of $s$. It may seem at a first glance, due to
the multiplication principle of combinatorics, that the
representation $K_{N}(s,p,k,l) = M_{\uparrow} M_{\downarrow}$ is
valid. This is, however, generally not the case. To obtain the
correct formula for $K_{N}(s,p,k,l)$ we note that for $p \geq 1$ the
variable $k$ can take only three values: $k=2p$, $k=2p-1$, and
$k=2p-2$. If $k=2p$ ($k=2p-2$) then both, the first and the last
domains in a chain belong to the down (up) type, and the previous
formula is valid. However, if $k=2p-1$ then those domains belong to
different types and, since the first domain can be either of up or
down type, we find for this case $K_{N}(s,p,k,l) = 2M_{\uparrow}
M_{\downarrow}$. Collecting the above results, we obtain
\begin{equation}
    K_{N}(s,p,k,l) = (1 + \delta_{k,2p-1})C_{s-l-1}^{p-2 +
    \Delta_{spkl}}C_{N-s-1}^{k-p}.
    \label{eq:number_K}
\end{equation}
Note that this representation of the function $K_{N}(s,p,k,l)$ is
valid if the values of its variables are compatible with each other.

\subsection{Three-variable joint probability functions}

By using Eqs.~(\ref{eq:joint4}) and (\ref{eq:number_K}), we next can
determine all the three-variable joint probability functions,
namely, $P_{N}(s,p,k)$, $P_{N}(p,k,l)$, $P_{N}(s,p,l)$, and
$P_{N}(s,k,l)$. In view of our purpose, however, i.e., for
determining the mentioned above one-variable probability functions,
we need only two of them, $P_{N}(s,p,k)$ and $P_{N}(p,k,l)$.
According to the common rule, to calculate the joint probability
function $P_{N}(s,p,k)$ we need to fix its variables and to perform
the summation of $P_{N}(s,p,k,l)$ over the admissible values of $l$.
Since $l=0$ (and $p=k=0$) if $s=0$, and $l=N$ (and $p=1$, $k=0$) if
$s=N$, and $1 \leq l \leq s-p+1$ if $1 \leq s \leq N-1$, we find
\begin{equation}
    P_{N}(s,p,k) = \left\{ \begin{array}{ll}
    \displaystyle W_{N}(0), \quad s=0,\; s=N, \\ [4pt]
    \displaystyle \tilde{P}_{N}(s,p,k), \quad 1 \leq s \leq N-1.
    \end{array}
    \right.
    \label{eq:joint3a}
\end{equation}
Here, we have used the conditions that $P_{N}(0,0,0,0)= W_{N}(0)$
and $P_{N}(N,1,0,N) = W_{N}(0)$, and introduced the notation
\begin{equation}
    \tilde{P}_{N}(s,p,k) = \sum_{l=1}^{s-p+1} P_{N}(s,p,k,l).
    \label{eq:joint3a_tilde}
\end{equation}

In order to evaluate the above sum, we evaluate first the sum $S_{1}
= \sum_{l=1}^{s-p+1}C_{s-l-1}^{p-2}$. If $p=1$ (i.e., $s\geq1$)
then, using the properties of binomial coefficients, we obtain
$S_{1} = \sum_{l=1}^{s}C_{s-l-1}^{-1} = C_{-1}^{-1} = 1$, and if
$p\geq2$ (i.e., $s\geq2$) then, using the relation \cite{PBM},
\begin{equation}
    \sum_{m=0}^{n}C_{c-m}^{b} = C_{c+1}^{b+1}-C_{c-n}^{b+1},
    \label{eq:sum1}
\end{equation}
being valid when the binomial coefficients exist, we have $S_{1} =
C_{s-1}^{p-1} - C_{p-2}^{p-1}$. Since $C_{p-2}^{p-1} = 0$ if
$p\geq2$ and $C_{s-1}^{p-1}=1$ if $p=1$, we conclude that the
formula $S_{1}=C_{s-1}^{p-1}$ holds for all $s\geq1$ (we recall that
$1 \leq p \leq s$). Therefore, substituting Eq.~(\ref{eq:number_K})
into Eq.~(\ref{eq:joint4}), from Eq.~(\ref{eq:joint3a_tilde}) we
obtain
\begin{equation}
    \tilde{P}_{N}(s,p,k) = (1 + \delta_{k,2p-1})W_{N}(k)
    C_{s-1}^{p-1}C_{N-s-1}^{k-p}.
    \label{eq:joint3a_tilde2}
\end{equation}
Although this formula has been derived for $1 \leq s \leq N-1$, its
right-hand side exists also for $s=0$ (when $p=k=0$) and $s=N$ (when
$p=1$ and $k=0$). Furthermore, since $\tilde{P}_{N}(0,0,0) =
\tilde{P}_{N}(N,1,0) = W_{N}(0)$, the desired  joint probability
function (\ref{eq:joint3a}) is given by the same expression, i.e.,
\begin{equation}
    P_{N}(s,p,k) = (1 + \delta_{k,2p-1})W_{N}(k)
    C_{s-1}^{p-1}C_{N-s-1}^{k-p}.
    \label{eq:joint3a_2}
\end{equation}

To evaluate $P_{N}(p,k,l)$, we need to find the admissible values of
$s$ for fixed $p$, $k$, and $l$. If $p=0$ then $s=0$ (and $p=l=0$),
if $p=1$ then $s=l$ for $k=1,2$ and $s=l=N$ for $k=0$, and if $2
\leq p \leq N/2$ then $p+l-1 \leq s \leq N-(k-p+1)$ (recall that
$k-p+1$ is the number of down domains in a chain). According to this
observation we get
\begin{equation}
    P_{N}(p,k,l) = \left\{ \begin{array}{ll}
    \displaystyle W_{N}(0), \quad p=0,\\ [4pt]
    \displaystyle P_{N}(l,1,k,l), \quad p=1,\\ [4pt]
    \displaystyle \tilde{P}_{N}(p,k,l), \quad 2 \leq p \leq N/2,
    \end{array}
    \right.
    \label{eq:joint3b}
\end{equation}
where
\begin{equation}
    P_{N}(l,1,k,l) = (1+\delta_{k,1})W_{N}(k)C_{N-l-1}^{k-1}
    \label{eq:joint3b_p=1}
\end{equation}
and
\begin{equation}
    \tilde{P}_{N}(p,k,l) = \sum_{s=p+l-1}^{N+p-k-1}P_{N}(s,p,k,l).
    \label{eq:joint3b_tilde}
\end{equation}
By use of the  relation \cite{PBM}
\begin{equation}
    \sum_{m=0}^{n}C_{c+m}^{c}C_{b-m}^{b-n} =
    C_{b+c+1}^{n}
    \label{eq:sum2}
\end{equation}
in evaluating
\begin{equation}
    \sum_{s=p+l-1}^{N+p-k-1}C_{s-l-1}^{p-2}C_{N-s-1}^{k-p}
    = C_{N-l-1}^{k-1},
    \label{eq:sum3}
\end{equation}
from Eq.~(\ref{eq:joint3b_tilde}) we obtain
\begin{equation}
    \tilde{P}_{N}(p,k,l) = (1 + \delta_{k,2p-1})W_{N}(k)
    C_{N-l-1}^{k-1}.
    \label{eq:joint3b_tilde2}
\end{equation}
Comparing this formula with (\ref{eq:joint3b_p=1}), we check that,
although Eq.~(\ref{eq:joint3b_tilde2}) is derived for $p\geq2$, it
remains also valid for $p=1$. Therefore, introducing the notation
$\Delta_{pkl} = \delta_{p,0}\delta_{k,0}\delta_{l,0}$, the result in
Eq.~(\ref{eq:joint3b}) can be represented in the appealing form
\begin{equation}
    P_{N}(p,k,l) = (1 + \delta_{k,2p-1})W_{N}(k)
    C_{N-l-1}^{k-1+\Delta_{pkl}}.
    \label{eq:joint3b_2}
\end{equation}

\subsection{Two-variable probability functions}

The four-variable joint probability function $P_{N}(s,p,k,l)$
generates six different two-variable joint probability functions.
But keeping in mind the one-variable probability functions, we
calculate only two of them, namely $P_{N}(p,k)$ and $P_{N}(k,l)$.
Because $P_{N}(p,k) = W_{N}(0)$ if $k=0$ and the parameter $s$
varies from $p$ to $N+p-k-1$ if $1 \leq k \leq N-1$, the former is
given by
\begin{equation}
    P_{N}(p,k) = \left\{ \begin{array}{ll}
    \displaystyle W_{N}(0), \quad k=0, \\ [4pt]
    \displaystyle \tilde{P}_{N}(p,k), \quad 1 \leq k \leq N-1,
    \end{array}
    \right.
    \label{eq:joint2a}
\end{equation}
where
\begin{equation}
    \tilde{P}_{N}(p,k) = \sum_{s=p}^{N+p-k-1} P_{N}(s,p,k).
    \label{eq:joint2a_tilde}
\end{equation}
Taking into account that, according to Eq.~(\ref{eq:sum2}), the
relation
\begin{equation}
    \sum_{s=p}^{N+p-k-1}C_{s-1}^{p-1}C_{N-s-1}^{k-p}
    = C_{N-1}^{k}
    \label{eq:sum4}
\end{equation}
holds (we used the condition $C_{n}^{n-m} = C_{n}^{m}$), we obtain
\begin{equation}
    P_{N}(p,k) = (1 + \delta_{k,2p-1})W_{N}(k)
    C_{N-1}^{k}.
    \label{eq:joint2a2}
\end{equation}

The same expression for $P_{N}(p,k)$ follows also from the joint
probability function (\ref{eq:joint3b_2}). Indeed, since for $1 \leq
k \leq N-1$ the parameter $l$ is varied from $1$ to $N-k$ (the
maximal value of $l$ corresponds to the case when all the remaining
$p-1$ up domains and all $k-p+1$ down domains consist of one spin),
we have
\begin{equation}
    \tilde{P}_{N}(p,k) = \sum_{l=1}^{N-k}P_{N}(s,p,l).
    \label{eq:joint2a_tilde2}
\end{equation}
Substituting Eq.~(\ref{eq:joint3b_2}) into
Eq.~(\ref{eq:joint2a_tilde2}), and using the formula $\sum_{l=1}^
{N-k} C_{N-l-1}^{k-1} = C_{N-1}^{k} - C_{k-1}^{k}$, which results
from Eq.~(\ref{eq:sum1}), and granting the conditions
$C_{k-1}^{k}=0$ ($k\neq0$) and $C_{N-1}^{0}=1$, we again arrive at
Eq.~(\ref{eq:joint2a2}).

In order to find $P_{N}(k,l)$ from Eq.~(\ref{eq:joint3b_2}), we
first notice that for fixed $k$ the parameter $p$ can take only one
or two values. More precisely, if $k$ is odd, i.e., $k=2h+1$ ($0
\leq h \leq N/2-1$), then the first and the last spins of a chain
belong to the different types. In this case, $p=h+1$, $1 \leq l \leq
N-2h-1$, and $P_{N}(k,l) = P_{N}(h+1,2h+1,l)$. On the contrary, if
$k$ is even, i.e., $k=2h$, then the first and the last spins belong
to the same type. In accordance with this, at $1 \leq l \leq N-2h$,
the parameter $p$ takes two values $p=h$ (if a chain begins and ends
by the down spins) and $p=h+1$ (if a chain begins and ends by the up
spins), and so $P_{N}(k,l) = P_{N}(h,2h,l) + P_{N}(h+1,2h,l)$.
Moreover, if $l=0$ ($p=k=0$) or $l=N$ ($p=1$, $k=0$) then
$P_{N}(k,l) = W_{N}(0)$. Combining these results yields
\begin{equation}
    P_{N}(k,l) = \left\{ \begin{array}{ll}
    \displaystyle W_{N}(0), \quad l=0,\, l=N, \\ [4pt]
    \displaystyle \tilde{P}_{N}(k,l), \quad 1 \leq l \leq N-1,
    \end{array}
    \right.
    \label{eq:joint2b}
\end{equation}
where
\begin{equation}
    \tilde{P}_{N}(k,l) = \sum_{p=[(k+1)/2]}^{[k/2]+1}P_{N}(p,k,l) \,
    \label{eq:joint2b_tilde}
\end{equation}
wherein $[x]$ denotes the integer part of $x$. Finally, taking into
consideration the following relation
\begin{equation}
    \sum_{p=[(k+1)/2]}^{[k/2]+1} (1+\delta_{k,2p-1})=2,
    \label{eq:sum5}
\end{equation}
we find from Eqs.~(\ref{eq:joint2b}), (\ref{eq:joint2b_tilde}), and
(\ref{eq:joint3b_2}) for the desired probability function the result
\begin{equation}
    P_{N}(k,l) = (2 - \delta_{l,0} - \delta_{l,N})W_{N}(k)
    C_{N-l-1}^{k-1+\Delta_{kl}}
    \label{eq:joint2b_2}
\end{equation}
($\Delta_{kl} = \delta_{k,0}\delta_{l,0}$).

\section{DISTRIBUTION OF DOMAIN WALLS}

According to Eqs.~(\ref{eq:joint2a2}) and (\ref{eq:sum5}), the
one-variable probability function
\begin{equation}
    P_{N}(k) = \sum_{p=[(k+1)/2]}^{[k/2]+1}P_{N}(p,k)\;,
    \label{eq:prob_k1}
\end{equation}
which characterizes the distribution of the number of domain walls
in a finite chain, is written as
\begin{equation}
    P_{N}(k) = 2W_{N}(k)C_{N-1}^{k}.
    \label{eq:prob_k2}
\end{equation}
The last formula reflects the fact that $P_{N}(k)$ is the overall
probability of all $2C_{N-1}^{k}$ spin configurations each of which
possesses $k$ domain walls and has the probability $W_{N}(k)$ (we
are grateful to an anonymous referee for this point). By using
Eqs.~(\ref{eq:partition1}) -- (\ref{eq:prob2}) and introducing the
designation $r=(1+e^{2\beta J})^{-1}$ ($0<r<1/2$), it can be recast
to read
\begin{equation}
    P_{N}(k) = C_{N-1}^{k}r^{k}(1-r)^{N-1-k}
    \label{eq:prob_k}
\end{equation}
($0 \leq k \leq N-1$).  This explicitly shows that a binomial
distribution for $k$ emerges.

The fact that the number of domain walls in an Ising chain is
binomially distributed has a simple interpretation. To demonstrate
this we first remind ourselves that the binomial distribution gives
the probability $C_{n}^{m} q^{m}(1-q)^{n-m}$ of $m$ successes in a
sequence of $n$ independent trials, called Bernoulli trials, each of
which has only one outcome, i.e., success with probability $q$ or
failure with probability $1-q$. In our case, we consider a chain as
a result of one-by-one addition of $N-1$ spins to the seed one. We
treat each addition as a trial whose outcome is either along or
opposite direction of the added spin with respect to the direction
of the nearest spin. If the added spin has the opposite direction
then the domain wall is formed and we call such outcome a success.
Hence, a chain of $N$ Ising spins with $k$ domain walls is
equivalent, in the above sense, to a sequence of $N-1$ Bernoulli
trials that have $k$ successes. Due to the exchange interaction, the
probability $q$ of success equals $e^{-\beta J}/(e^{-\beta J} +
e^{\beta J}) = r$, and so the probability that a chain has exactly
$k$ domain walls is indeed given by Eq.~(\ref{eq:prob_k}).

The probability function (\ref{eq:prob_k}) is properly normalized,
i.e., $\sum_{k=0}^{N-1} P_{N}(k) = 1$, and, in accordance with
well-known properties of the binomial distribution \cite{F}, the
average $\langle k \rangle = \sum_{k=1}^{N-1} kP_{N}(k)$ and the
variance $\sigma_{k}^{2} \equiv \langle k^{2} \rangle - \langle k
\rangle^{2} = \sum_{k=1}^{N-1} k^{2}P_{N}(k) - \langle k
\rangle^{2}$ of the number of domain walls in a chain assume the
form
\begin{equation}
    \langle k \rangle = (N-1)r, \quad \sigma_{k}^{2} = (N-1)(1-r)r.
    \label{eq:mean_var_k}
\end{equation}
For $\beta J \ll 1$, we obtain $\langle k \rangle = (N - 1)(1 -
\beta J)/2$ and $\sigma_{k}^{2} = (N - 1)(1 - \beta^{2}J^{2})/4$
with linear and quadratic accuracy in $\beta J$, respectively. The
relation $\lim_{\beta J \to 0} \langle k \rangle = (N - 1)/2$ makes
explicit that in the high-temperature case, which is characterized
by the condition $r=1/2$, approximately one domain wall falls on two
spins, implying that each domain contains, on average, two spins
(see also Sec.~VI).

The increase of $\beta J$ leads to the decrease of $r$ and
Eq.~(\ref{eq:mean_var_k}) yields $\langle k \rangle \approx
\sigma_{k}^{2} \approx (N-1)e^{-2\beta J}$ for large enough values
of $\beta J$. If $2\beta J \gg \ln N$ then the condition $\langle k
\rangle \ll 1$ holds, which indicates that in this case the spin
configurations $\{ \sigma_{i} = 1 \}$ and $\{ \sigma_{i} = -1 \}$
play the main role in determining the chain properties. Let
$\tau_{\mathrm{tr}}$ and $\tau_{\mathrm {m}}$ be the transition time
between these states, i.e., the average time during which a chain
passes from the state $\{ \sigma_{i} = 1 \}$ to the state $\{
\sigma_{i} = - 1 \}$, and the measurement time, i.e., the total time
necessary to perform a measurement of the magnetization,
respectively. Then, if $2\beta J \gg \ln N$ and $\tau_{\mathrm{tr}}
\gg \tau_{\mathrm {m}}$, a chain possesses a spontaneous
magnetization. In other words, these conditions form a criterion of
the ferromagnetic-like behavior of a finite Ising chain. Notice that
in the thermodynamic limit ($N \to \infty$) the second condition
holds always, see just below, while the first condition holds only
if $T=0$. Therefore, in full agreement with \cite{Ising}, the
long-range ferromagnetic order in an infinite chain occurs only at
zero temperature.

In order to estimate the dependence of $\tau_{\mathrm{tr}}$ on $N$,
it is necessary to go beyond the Ising model. To this end, we
consider the Ising spins as the classical Heisenberg spins with
large uniaxial anisotropy and use the Arrhenius-Neel law
\cite{N,RMP90}. According to it, the average time $\tau$ between the
spin reversals can be evaluated as $\tau \sim \tau_{0} e^{\beta
\Delta U}$, where $\tau_{0}$ is the spin precession time, and
$\Delta U (\gg \beta^{-1})$ is the height of the potential barrier
between two equilibrium directions of each spin. Since a chain in a
state with $\langle k \rangle \ll 1$ can be associated with a single
enlarged spin for which the potential barrier height is given by
$N\Delta U$, we find that the transition time $\tau_{ \mathrm{tr}}$
exponentially grows with $N$: $\tau_{\mathrm{tr}} \sim \tau
e^{(N-1)\beta \Delta U}$. Note also that because $\beta J \to
\infty$ and $\tau_{\mathrm{tr}} \to \infty$ as $T \to 0$, there is
always a temperature interval where a finite Ising chain exhibits
the ferromagnetic-like behavior.

We briefly discuss here also the problem of domain walls
distribution in an infinite chain. As is well known \cite{S}, the
binomial distribution has no unique asymptotic as the number of
Bernoulli trials tends to infinity. However, since in our case the
parameter $r$ does not depend on $N$, the probability function
$P_{N}(k)$ does have it. To characterize $P_{N}(k)$ as $N \to
\infty$, we assume in Eq.~(\ref{eq:prob_k}) that $k = \langle k
\rangle + \sigma_{k}z$ and define the probability function $P_{k}(z)
= \lim_{N \to \infty}\sigma_{k}P_{N} (\langle k \rangle +
\sigma_{k}z)$ of the parameter $z$. Applying a local limit theorem
\cite{S}, we immediately find that $P_{k}(z)$ has the standard
Gaussian distribution: $P_{k}(z) = (2\pi)^{-1/2} e^{-z^{2}/2}$.

\section{DISTRIBUTION OF UP DOMAINS}

To derive the one-variable probability function $P_{N}(p)$ that
describes the distribution of the number of up domains in a finite
chain, we again proceed from the joint probability function
$P_{N}(p,k)$. A simple consideration shows that $k=0$ if $p=0$,
$k=2p-i$ ($i=0,1,2$) if $1 \leq p \leq N/2-1$, and $k=2p-i$
($i=1,2$) if $p = N/2$. Hence, for fixed $p$ the parameter $k$ is
varied from $2p-2 + 2\delta_{p,0}$ to $2p-\delta_{p,N/2}$, and
$P_{N}(p)$ is given by
\begin{equation}
    P_{N}(p) = \sum_{k=2p-2+2\delta_{p,0}}^{2p-\delta_{p,N/2}}
    P_{N}(p,k).
    \label{eq:prob_p1}
\end{equation}
Substituting Eq.~(\ref{eq:joint2a2}) into this formula and taking
into account the properties of binomial coefficients, this
probability distribution is obtained as

\begin{equation}
    P_{N}(p) = \sum_{n=0}^{2} (1 + \delta_{1,n})W_{N}(2p-n)
    C_{N-1}^{2p-n}.
    \label{eq:prob_p2}
\end{equation}
Finally, using Eqs.~(\ref{eq:partition1}) -- (\ref{eq:prob2}), from
Eq.~(\ref{eq:prob_p2}) we find the desired probability function in
the form
\begin{equation}
    P_{N}(p) = \frac{1}{2}\sum_{n=0}^{2}(1+\delta_{1,n})
    C_{N-1}^{2p-n}r^{2p-n}(1-r)^{N-1-2p+n}
    \label{eq:prob_p}
\end{equation}
($0 \leq p \leq N/2$). This distribution, due to its formal
closeness to the ordinary binomial distribution, will be termed the
\textit{modified binomial distribution}.

Taking into consideration the results for the finite series
\cite{PBM}
\begin{equation}
    \sum_{k=0}^{[n/2]}C_{n}^{2k}x^{k} = \frac{(1+\sqrt{x})^{n}
    + (1-\sqrt{x})^{n}}{2},
    \label{eq:sum6}
\end{equation}
\begin{equation}
    \sum_{k=0}^{[(n-1)/2]}C_{n}^{2k+1}x^{k} = \frac{(1+\sqrt{x})^{n}
    - (1-\sqrt{x})^{n}}{2\sqrt{x}}
    \label{eq:sum7}
\end{equation}
and the properties of the binomial coefficients, one readily finds
that the quantities
\begin{equation}
    I_{n}^{m} = \sum_{p=0}^{N/2}C_{N-1-m}^{2p-n-m}r^{2p-n}(1 -
    r)^{N-1-2p+n}
    \label{eq:def_I}
\end{equation}
($n,m=0,1,2$) can be represented in the form
\begin{equation}
    I_{n}^{m} = \frac{r^{m}}{2} \left[1 + (-1)^{n+m}(1 - 2r)^
    {N-1-m}\right].
    \label{eq:res_I}
\end{equation}
With these results, it follows that the modified binomial
distribution is also properly normalized, namely,
\begin{equation}
    \sum_{p=0}^{N/2}P_{N}(p) = \frac{1}{2}\sum_{n=0}^{2}(1 +
    \delta_{1,n})I_{n}^{0} = 1.
    \label{eq:norma_p}
\end{equation}

The average of the number of up domains in a chain is defined as
$\langle p \rangle = \sum_{p=0}^{N/2}p \,P_{N}(p)$. Using the
probability function (\ref{eq:prob_p}) and the identity
\begin{equation}
    2p\,C_{N-1}^{2p-n} = nC_{N-1}^{2p-n} + (N-1)C_{N-2}^{2p-n-1},
    \label{eq:identity1}
\end{equation}
which can be verified directly, we find
\begin{equation}
    \langle p \rangle =\frac{1}{4}\sum_{n=0}^{2}(1 + \delta_{1,n})
    [nI_{n}^{0}+(N-1)I_{n}^{1}],
    \label{eq:def_<p>}
\end{equation}
and substituting Eq.~(\ref{eq:res_I}) into Eq.~(\ref{eq:def_<p>}) we
obtain
\begin{equation}
    \langle p \rangle = \frac{1}{2} + \frac{1}{2}(N-1)r.
    \label{eq:mean_p}
\end{equation}
It may seem strange at a first glance that $\langle p \rangle = 1/2$
in the low-temperature limit ($r \to 0$), but in compliance with
Eq.~(\ref{eq:prob_p}) at $r \to 0$ only two states of a chain,
namely $\{ \sigma_{i} = -1\}$ ($p=0$) and $\{ \sigma_{i} = 1\}$
($p=1$), have nonzero probabilities and they are equal to $1/2$.
Notice also that, according to Eqs.~(\ref{eq:mean_var_k}) and
(\ref{eq:mean_p}), the general condition $2\langle p \rangle = 1 +
\langle k \rangle$ always holds.

To find the variance $\sigma_{p}^{2} = \langle p^{2} \rangle -
\langle p \rangle^{2}$ of the number of up domains in a chain, we
first calculate the second moment $\langle p^{2} \rangle =
\sum_{p=0}^{N/2}p^{2}P_{N}(p)$. By applying the identity
{\setlength\arraycolsep{2pt}
\begin{eqnarray}
    4p^{2}C_{N-1}^{2p-n} &=& n^{2}C_{N-1}^{2p-n} +
    (2n+1)(N-1)C_{N-2}^{2p-n-1}
    \nonumber\\[3pt]
    && +\, (N-1)(N-2)C_{N-3}^{2p-n-2}
    \label{eq:identity2}
\end{eqnarray}}(note that last term equals zero at $N=2$) and the
notation (\ref{eq:def_I}), this quantity can be expressed as
{\setlength\arraycolsep{2pt}
\begin{eqnarray}
    \langle p^{2} \rangle &=& \frac{1}{8}\sum_{n=0}^{2}(1
    + \delta_{1,n})[n^{2}I_{n}^{0}+(2n+1)(N-1)I_{n}^{1}
    \nonumber\\[3pt]
    && +\, (N-1)(N-2)I_{n}^{2}].
    \label{eq:moment_p}
\end{eqnarray}}Inserting Eq.~(\ref{eq:res_I}) into this formula and
performing straightforward calculations yields
{\setlength\arraycolsep{2pt}
\begin{eqnarray}
    \langle p^{2} \rangle &=& \frac{3}{8} +
    \frac{3}{4}(N-1)r + \frac{1}{4}(N-1)(N-2)r^{2}
    \nonumber\\[3pt]
    && +\, \frac{1}{8}(1-2r)^{N-1}\;.
    \label{eq:moment_p2}
\end{eqnarray}}Therefore, using the definition of the variance
$\sigma_{p}^{2}$, we find
\begin{equation}
    \sigma_{p}^{2} = \frac{1}{8} + \frac{1}{4}(N-1)(1-r)r +
    \frac{1}{8}(1-2r)^{N-1}.
    \label{eq:var_p}
\end{equation}
The fact that $\langle p^{2} \rangle \to 1/2$ ($\sigma_{p}^{2} \to
1/4$) as $r \to 0$ has the same interpretation as the
low-temperature behavior of $\langle p \rangle$ given above.

In conclusion, we note that if $p = \langle p \rangle + \sigma_{p}z$
as $N \to \infty$ then the parameter $z$ again possesses a standard
Gaussian distribution (see Appendix~A).

\section{DISTRIBUTION OF DOMAIN LENGTHS}

To find the probability function of domain lengths, $P_{N}(l)$, we
proceed from the joint probability function (\ref{eq:joint2b_2}).
Since for $1 \leq l \leq N-1$ the number of domain walls $k$ can
vary from $1$ to $N-l$ and $k=0$ if $l=0$ or $l=N$, this probability
function can be written as
\begin{equation}
    P_{N}(l) = \left\{ \begin{array}{ll}
    W_{N}(0), \quad l=0,\, l=N, \\ [4pt]
    \tilde{P}_{N}(l), \quad 1 \leq l \leq N-1,
    \end{array}
    \right.
    \label{eq:prob_l}
\end{equation}
where
\begin{equation}
    \tilde{P}_{N}(l) = \sum_{k=1}^{N-l}P_{N}(k,l).
    \label{eq:prob_l_tilde}
\end{equation}
In virtue of this, taking into account that $W_{N}(0) =
(1-r)^{N-1}/2$ and using the standard series $\sum_{k=0}^{n}
C_{n}^{k}x^{k} = (1+x)^{n}$ which permits us to reduce
Eq.~(\ref{eq:prob_l_tilde}) into the form $\tilde{P}_{N}(l) =
r(1-r)^{l-1}$, we obtain
\begin{equation}
    P_{N}(l) = \left\{ \begin{array}{ll}
    (1-r)^{N-1}/2, \quad l=0,\, l=N, \\ [4pt]
    r(1-r)^{l-1}, \quad 1 \leq l \leq N-1.
    \end{array}
    \right.
    \label{eq:prob_lb}
\end{equation}
It is not difficult to verify, using the well-known relation,
$\sum_{k=0}^{n}x^{k} = (1-x^{n+1})/(1-x)$, that this distribution,
which we term the \textit{finite geometric distribution}, is
normalized, i.e., $\sum_{l=0}^{N} P_{N}(l) = 1$. Note also that in
the limit of an infinite chain the domain lengths distribution
(\ref{eq:prob_lb}) is reduced to the geometric distribution,
$P_{\infty}(l) = r(1-r)^{l-1}$ ($l=1,2, \ldots$), whose mean and
variance are $1/r$ and $(1-r)/r^{2}$, respectively.

By applying the standard series, $\sum_{k=0}^{n}kx^{k} = [x +
(nx-n-1) x^{n+1}] /(1-x)^{2}$, the average length of an up domain,
$\langle l \rangle = \sum_{l=0}^ {N}lP_{N}(l)$, can be represented
in the form
\begin{equation}
    \langle l \rangle = \frac{2-Nr(1-r)^{N-1}-2(1-r)^{N}}{2r}.
    \label{eq:mean_l}
\end{equation}
An alternative derivation of this result is presented in the
Appendix~B. According to this expression, we find $\lim_{N \to
\infty} \langle l \rangle = 1/r$, $\lim_{r \to 0} \langle l \rangle
= N/2$, and $\lim_{r \to 1/2} \langle l \rangle = 2 - (N+2)2^{-N}$.
The last condition shows that in the high-temperature limit the
average number of spins that form one up domain in a long chain is
approximately equal to 2.

All other moments of the finite geometric distribution
(\ref{eq:prob_lb}) are also calculated exactly. In particular, the
variance of domain lengths, $\sigma_{l}^{2} = \langle l^{2} \rangle
- \langle l \rangle^{2}$, is given by {\setlength\arraycolsep{2pt}
\begin{eqnarray}
    \sigma_{l}^{2} &=& (1-r)/r^{2} - (N/2)^{2}[2+(1-r)^{N-1}]
    (1-r)^{N-1}
    \nonumber\\[3pt]
    && -\, N[1-2r+(1-r)^{N}](1-r)^{N-1}/r
    \nonumber\\[3pt]
    && +\, [r-(1-r)^{N}](1-r)^{N}/r^{2}.
    \label{eq:var_l}
\end{eqnarray}}With this result we immediately obtain
$\sigma_{l}^{2} \to (1-r)/r^{2}$ as $N \to \infty$, $\sigma_{l}^{2}
\to N^{2}/4$ as $r \to 0$, and $\sigma_{l}^{2} \to 2 -
(N^{2}-2)2^{-N} -(N+2)^{2}2^{-2N}$ as $r \to 1/2$.

To gain more insight into the domain statistics, we also introduce
the probability function $P_{N}^{+}(l)$ that describes the domain
lengths distribution in assemblies of Ising chains each of which
contains at least one up domain of nonzero length. In other words,
we assume that the parameter $l$ can vary from $1$ to $N$. In this
case, in contrast with Eq.~(\ref{eq:joint2b_2}), the joint
probability function $P_{N}^{+}(k,l)$ that a chain contains $k$
domain walls and the first up domain contains $l$ spins is written
as
\begin{equation}
    P_{N}^{+}(k,l) = (2 - \delta_{l,N})W_{N}^{+}(k)C_{N-l-1}^{k-1},
    \label{eq:prob+_l}
\end{equation}
where $W_{N}^{+}(k) = e^{-\beta E_{N}(k)}/Z_{N}^{+}$ and $Z_{N}^{+}
= Z_{N} - e^{\beta J(N-1)}$ is the partition function for such a
chain. Therefore, the probability function $P_{N}^{+}(l)$, which is
defined as $P_{N}^{+}(l) = \sum_{k=1 - \delta_{l,N}}^{N-l}
P_{N}^{+}(k,l)$, assumes the form
\begin{equation}
    P_{N}^{+}(l) = \frac{2}{2-(1-r)^{N-1}}\left\{
    \begin{array}{ll}
    (1-r)^{N-1}/2,  \quad l=N, \\[4pt]
    r(1-r)^{l-1}, \quad 1 \leq l \leq N \!-\! 1.
    \end{array}
    \right.
    \label{eq:prob_l2}
\end{equation}

One can again verify  that the normalization condition $\sum_{l=1}
^{N}P_{N}^{+}(l) = 1$ holds, and that the average length of an up
domain, $\langle l \rangle^{+} = \sum_{l=1} ^{N}lP_{N}^{+}(l)$, can
be represented as
\begin{equation}
    \langle l \rangle^{+} = \frac{2-Nr(1-r)^{N-1}-2(1-r)^{N}}
    {2r-r(1-r)^{N-1}}.
    \label{eq:mean_l2}
\end{equation}
As $N \to \infty$, the averages $\langle l \rangle^{+}$ and $\langle
l \rangle$ tend to the same limit, but their low- and
high-temperature limits are different: $\lim_{r \to 0}\langle l
\rangle^{+} = N$ and $\lim_{r \to 1/2} \langle l \rangle^{+} = 2 -
N/(2^{N}-1)$. The former confirms the possibility of the existence
of the ferromagnetic-like state in a finite Ising chain at low
temperatures. We note in this context that the condition $2\beta J
\gg \ln N$, which we introduced in Sec.~IV, is equivalent to the
condition $\xi \gg N$, where $\xi$ is the spin-spin correlation
length. According to \cite{Domb}, in the case of Ising chains with
only exchange interaction and free boundary conditions the
correlation length is given by the exact formula $\xi =
-\ln^{-1}(1-2r)$. Since at low temperatures the asymptotic relations
$\xi \sim 1/2r$ and $r \sim e^{-2\beta J}$ hold, the condition $\xi
\gg N$ is actually reduced to $2\beta J \gg \ln N$.

\section{CONCLUSIONS}

We have determined the domain statistics in a finite chain of Ising
spins that interact only through the exchange interaction. For a
chain with an even number of spins and free boundary conditions, we
have calculated, via a combinatorial approach, the joint probability
function of four random variables (namely, the number of up spins,
the number of up domains, the number of domain walls, and the number
of spins in the first up-domain) that thoroughly describe the domain
statistics in a chain. Starting out from this result, we derived the
probability distribution functions for the number of domain walls,
number of up domains, and number of spins in an up domain. The first
corresponds to the binomial distribution, the second to the modified
binomial distribution, and the third to the finite geometric
distribution. For each of them, we have calculated the corresponding
thermal average and variance, have analyzed the cases of low and
high temperatures, and, as well, have considered the thermodynamic
limit.

In addition, we have derived a criterion that a finite Ising chain
exhibits the ferromagnetic-like behavior. According to it, the
transition time between the fully magnetized chain states must
exceed the measuring time, and the average number of domain walls
must be much less than 1. These conditions hold, i.e., a finite
Ising chain does display a ferromagnetic-like order on the measuring
time scale, if the temperature is sufficiently small.

\section*{ACKNOWLEDGMENTS}

S.I.D. acknowledges the support by the European Union under Contract
No. NMP4-CT-2004-013545, and P.H. the support by the Deutsche
Forschungsgemeinschaft, grant HA 1517/13-4.

\appendix

\section{DISTRIBUTION OF DOMAINS IN AN INFINITE CHAIN}

To find the probability function of the parameter $z$, $P_{p}(z) =
\lim_{N \to \infty} \sigma_{p}P_{N}(\langle p \rangle +
\sigma_{p}z)$, first we represent the binomial coefficients in
Eq.~(\ref{eq:prob_p}) as
\begin{equation}
    C_{a}^{b} = \frac{\Gamma(a+1)}{\Gamma(b+1)\Gamma(a-b+1)}
    \label{eq:def_binom}
\end{equation}
($\Gamma(x)$ is the gamma function) and use the Stirling formula
\cite{AS}
\begin{equation}
    \Gamma(x) \sim \sqrt{2\pi}\,e^{-x}x^{x-1/2} \quad (x \to
    \infty).
    \label{eq:stirling}
\end{equation}
This yields
\begin{equation}
    C_{N-1}^{2\langle p \rangle+2\sigma_{p}z-n} \!\sim\!
    \frac{eR_{n}}{\sqrt{2\pi N}(1\!-r)^{N}}
    \Bigl( \frac{1\!-r}{r} \Bigr)^{Nr+2\sigma_{p}z+3/2-r-n}
    \label{eq:asymp1}
\end{equation}
and
\begin{eqnarray}
    R_{n} \!&\sim&\! \Bigl(1\! - \frac{2\sigma_{p}z + 1 - r - n}
    {N(1-r)}\Bigr)^{\!-N(1-r) + 2\sigma_{p}z + 3/2 - r - n}
    \nonumber\\[3pt]
    && \!\times \Bigl(1\! + \frac{2\sigma_{p}z + 2 - r - n}
    {Nr}\Bigr)^{\!-Nr - 2\sigma_{p}z - 3/2 + r + n}
    \nonumber\\
    \label{eq:asymp2}
\end{eqnarray}
as $N \to \infty$, and so
\begin{equation}
    P_{p}(z) = \lim_{N \to \infty} \frac{e\sigma_{p}}
    {2\sqrt{2\pi r(1-r)N}}\sum_{n=0}^{2}(1+\delta_{1,n})R_{n}.
    \label{eq:prob_P}
\end{equation}
Finally, taking into account that $\sigma_{p}^{2}/N \to r(1-r)/4$
and $\ln R_{n} \to -z^{2}/2 -1$ as $N \to \infty$, we indeed find
that $P_{p}(z) = (2\pi)^{-1/2} e^{-z^{2}/2}$.

\section{ALTERNATIVE DERIVATION OF EQUATION~(\ref{eq:mean_l})}

Using the joint probability function (\ref{eq:joint3a_2}), we can
also represent $\langle l \rangle$ in the following form:
\begin{equation}
    \langle l \rangle = \sum_{p=1}^{N/2}\frac{1}{p}\,\sum_{k=2p-2}^
    {2p-\delta_{p,N/2}}\,\sum_{s=p}^{N+p-k-1}sP_{N}(s,p,k).
    \label{eq:def_<l>}
\end{equation}
Since $sC_{s-1}^{p-1} = p\,C_{s}^{p}$, $W_{N}(k) = r^{k}(1-r)
^{N-k-1}/2$, and according to the result for the series
(\ref{eq:sum2})
\begin{equation}
    \sum_{s=p}^{N+p-k-1}s\,C_{s-1}^{p-1}C_{N-s-1}^{k-p} =
    p\,C_{N}^{k+1},
    \label{eq:sum8}
\end{equation}
Eq.~(\ref{eq:def_<l>}) can be rewritten as
\begin{equation}
    \langle l \rangle = \frac{(1-r)^{N-1}}{2}\sum_{n=0}^{2}
    (1+\delta_{1,n})Y_{n},
    \label{eq:mean_lb}
\end{equation}
where
\begin{equation}
    Y_{n} = \sum_{p=1}^{N/2}\Bigl(\frac{r}{1-r}\Bigr)
    ^{2p-n}C_{N}^{2p+1-n}.
    \label{eq:def_Y}
\end{equation}
Upon calculating these quantities with the help of the  series
(\ref{eq:sum6}) and (\ref{eq:sum7}),
\begin{equation}
    Y_{n} = \frac{1\!-r}{2r}\Bigl[\Bigl(\frac{1}{1\!-r}
    \Bigr)^{N} - (-1)^{n}\Bigl(\frac{1\!-2r}{1\!-r}
    \Bigr)^{N}-2\delta_{1,n}\Bigr] - N\delta_{0,n}
    \label{eq:res_Y}
\end{equation}
($n=0,1,2$), and substituting the corresponding expressions into
Eq.~(\ref{eq:mean_lb}), we again obtain Eq.~(\ref{eq:mean_l}).

\end{document}